\documentclass[letter,twocolumn]{jpsj3}
\usepackage{amsmath,amssymb}
\usepackage{graphicx}
\usepackage{bm}
\usepackage{algorithm}
\usepackage{algpseudocode}
\usepackage[usenames]{color}

\title{An exact algorithm exhibiting RS-RSB/easy-hard correspondence \\
for the maximum independent set problem}
\date{\today}

\author{\name{Jun Takahashi$^1$}\thanks{E-mail: jt@huku.c.u-tokyo.ac.jp}, \name{Satoshi Takabe $^2$}\thanks{E-mail: s{\_}takabe@nitech.ac.jp} and \name{Koji Hukushima}$^1$\thanks{E-mail: hukusima@phys.c.u-tokyo.ac.jp}}
\inst{$^1$ Graduate School of Arts and Sciences, The University of Tokyo, Meguro, Tokyo 153-8902, Japan\\
$^2$ Department of Computer Science, Nagoya Institute of Technology, Gokiso-cho, Showa-ku, Nagoya, Aichi, 466-8555, Japan}

\abst{
 A recently proposed exact
algorithm
for the maximum independent set problem
is analyzed.
The typical running time is improved exponentially in some parameter regions 
compared to 
simple binary search.
The algorithm also overcomes the core transition point, where the conventional 
leaf removal algorithm fails, and works up to the replica symmetry breaking (RSB) transition point.
This suggests that a leaf removal core itself is not enough for typical hardness in the random maximum independent set problem,
providing further evidence for RSB being the obstacle for algorithms in general.
}
\kword{maximum independent set, average case complexity, random graphs, scale free networks, replica symmetry breaking, dynamic programming, leaf removal}

\begin{document}

\maketitle

Studies in the field of statistical physics dealing with spin glasses have developed in the 1970s 
resulting in fruitful applications to a number of areas \cite{SGTandB}.
Computer science is one of them, 
and mean-field spin-glass theory has provided excellent tools 
for analyzing random combinatorial optimization problems \cite{IPC}.
The central feature provided by the statistical physics frame work is,
that many combinatorial optimization problems \cite{2pluspSAT, coloring, MIS} 
show a phase transition called replica symmetry breaking (RSB).
Since the replica symmetric (RS) phase has a smooth and connected solution space,
and RSB phases correspond to rugged, non-ergodic solution space,
it is generally thought that an RS/RSB phase transition should make the 
random optimization problem hard.
In this paper, we will refer to this perspective as the {\it RS-RSB/easy-hard correspondence}.
Some of this RS-RSB/easy-hard correspondence has been made rigorous in few specific models\cite{Amin-coja-oghlan, Cocco-Monasson},
which show that a certain algorithm takes only polynomial time to run in the RS phase,
and takes exponentially long time in the RSB phase.
However, extensions for general cases seem to be difficult, and
whether if the correspondence generally holds or not remains as an open problem. 
Indeed, the most naive form of the correspondence is violated when we consider 
the XORSAT problem \cite{XORSAT-math, XORSAT-phys}, since although it exhibits an RSB transition,
it always has a polynomial time algorithm.
In order to restore the correspondence, it is likely that an analysis which takes the algorithmic aspect into account is needed. 
Conversely to the XORSAT case, some natural algorithms can fail even in the RS region,
as we will discuss in the following.
Thus, a good algorithm and a suited statistical physics analysis is needed in general to explore the validity of the correspondence.

In this work we will focus on the randomized maximum independent set (MIS) problem,
in which a previously discussed algorithm fails to achieve the correspondence.
We examine a recently proposed exact 
algorithm\cite{UniqueMIS},  which 
solves the MIS problem exponentially faster compared to the previous algorithm in some parameter region,
resulting in polynomial-time computation up to the RS/RSB transition point.
The MIS problem could be formalized as follow. 
The input of the problem is a simple graph $G=(V,E)$, where $V$ is a vertex set and $E\subset V^2$ is an edge set.
We denote the number of the vertices $N:=|V|$, and label each vertex by $i\in \{1,2,\ldots,N\}$.
A subset of the vertices $I\in V$ such that no two vertices $i,j$ in $I$ are connected in graph $G$ is called an {\it independent set}.
The task of the problem is to find the maximum possible independent set for a given graph.
We can think of a binary variable $x_i \in \{0,1\}$ assigned to each vertex $i$. 
Then, the MIS problem could be thought of as an problem of finding the ground state
of a Hamiltonian with 
\begin{equation}
H=-\sum_{i} x_i + \alpha\sum_{(i,j)\in E} x_i x_j \label{eq:Hamiltonian}
\end{equation}
where $\alpha >1 $ is a constant.
In terms of computational complexity theory,
the MIS problem is NP-hard.
This implies that there is no algorithm that exactly solves the MIS problem
with a polynomial upper bound on the running time,
as long as if P$\neq$NP.
To consider random instances for the MIS problem,
a probability distribution over the input graphs is introduced.
This corresponds to random graphs, which we will discuss below in detail,
focusing on two types of random graph ensembles.

For Erd\"{o}s-R\'{e}nyi random graphs,
previous study shows that a phase transition occurs when the average degree $c$ crosses 
the Napier's constant $e\simeq 2.718$ \cite{HartmannWeigt}.
The transition could be understood in
two different ways.
Physically, it is a
phase transition from the RS phase to the full RSB phase \cite{ZhouVC}.
From the algorithmic point of view, the RS/RSB transition point is where the Leaf Removal (LR) algorithm \cite{Karp-Sipser} seizes to work,
and leaves a so-called LR-core with $O(N)$ vertices undecided \cite{LeafRemoval}.
We can simply express this as $c_{\mathrm{RSB}}= c_{\mathrm{LR}}=e$.
Furthermore, it is also shown that linear relaxation, which is another type of algorithmic approach, fails at this point as well \cite{TakabeHukushima}.
The fact that the two algorithms starts to fail 
at the RS/RSB transition point, can be seen as a concrete example 
of the RS-RSB/easy-hard correspondence.

However, the situation differs when other graph ensembles are considered.
Below, we will focus on a configuration model \cite{ConfigModel} 
with a power-law degree distribution
\begin{equation}
p_k=\begin{cases}
0 & (k<m)\\
\frac{2(1-p)}{m+2} & (k=m)\\
\frac{2m(m+1)}{k(k+1)(k+2)}(1-p)+\frac{2(m+1)(m+2)}{k(k+1)(k+2)}p & (k>m),
	\end{cases}
	\label{eq:pk}
\end{equation}
where $c$ is the average degree and $m:=\lfloor c/2 \rfloor$ and $p:=c/2-m$.
This could be seen as a generalization of the 
Barab\'{a}si-Albert model \cite{BAmodel} by linear combination,
but without any degree correlation. 
We will call this ensemble the CBA random graph model.
The absence of degree correlation enables statistical mechanics analysis, and
it is known that there is
 a RS/RSB transition at $c_{\mathrm{RSB}}\simeq 5.239$ \cite{LRfailsforSF}.
However,
all the vertices have degree $\geq 2$ for $c \geq c_{\mathrm{LR}}=4$,
where the entire graph becomes the LR core.
Thus, $c_{\mathrm{LR}}\lneq c_{\mathrm{RSB}}$ for this graph ensemble,
meaning that the easy/hard transition for the LR algorithm does not correspond to 
the RS/RSB transition.
The most natural way to interpret the disagreement
is that the 
LR algorithm by itself is too poor and naive to illustrate the correspondence
for CBA random graphs.
We thus introduce a natural extension of the LR algorithm, achieving the correspondence.

\begin{figure}[h]
\includegraphics[width=8.5cm]{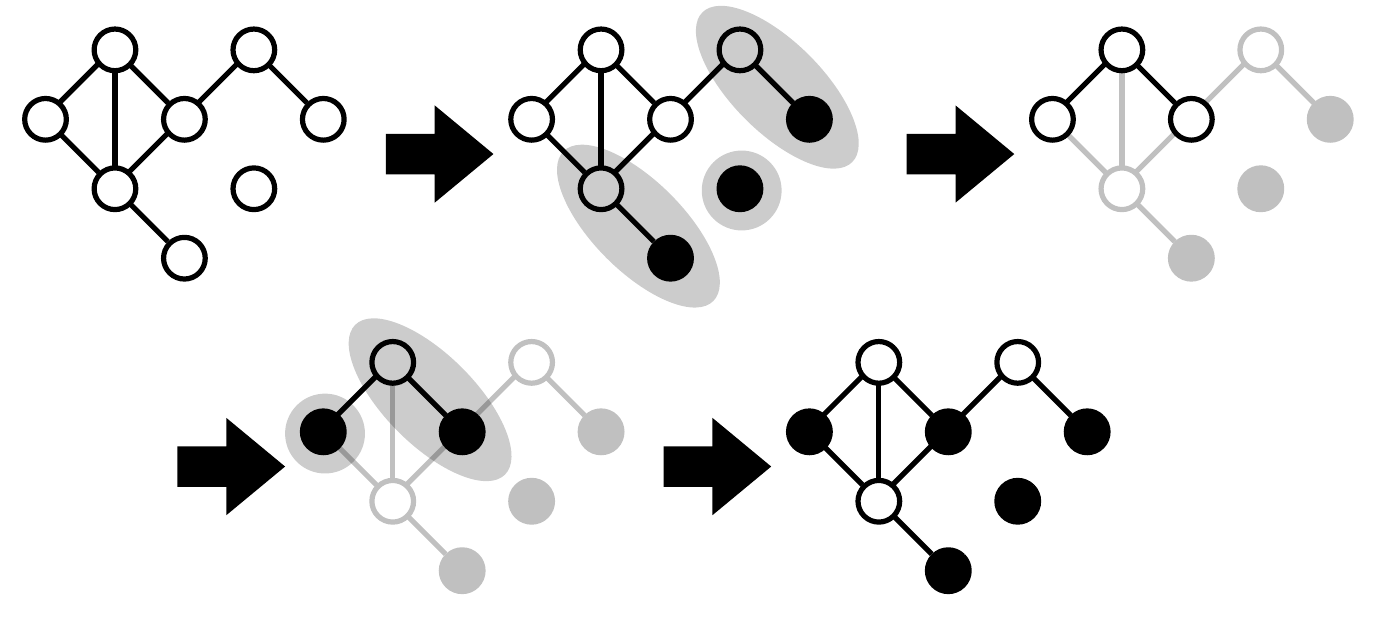}
\caption{\label{Fig:LR} 
A schematic diagram for the leaf removal algorithm.
Black colored vertices are the ones decided to be {\tt included} in the independent set.
The shadowed vertices are the parts considered as the ``leaves",
and are removed from the graph, becoming pale.
 }
\end{figure}

We will first explain the LR algorithm in detail.
The LR algorithm decides which vertices to include in the independent set 
with a guarantee that at least one of the MIS indeed includes those vertices.
Any vertex $v$ with degree 0 is trivially included in the MIS,
so it is labeled as {\tt included} and is removed from the graph.
Any vertex $v$ with degree 1 is also labeled {\tt included} and is removed as well.
This is because there exists at least one pattern for the MIS  
which includes $v$, since 
either of 
vertex $v$ or the 
neighboring vertex $w$ must be included 
in order to achieve the maximum,
and $v$ could be chosen without harm.
When a vertex $v$ with degree 1 is removed, the 
neighboring vertex $w$
will be removed as well,
since if $v$ is in the MIS, $w$ cannot be.
This will be expressed as $w$ 
being labeled {\tt excluded}.
The procedure is continued until there are no longer any vertices with degree less than 2.
Fig. \ref{Fig:LR} shows a schematic diagram for this algorithm.

Intuitively, the LR algorithm removes the {\it leaves} from the graph,
which are actually nonessential to the intrinsic hardness of the MIS problem.
By removing the leaves, new leaves may emerge by the reduction of edges,
and if most of the graph turns out to become a leaf,
the LR algorithm is successful.
The remaining vertices when the algorithm stops is called the LR-core,
which is a sub-graph of the original input graph $G$, 
only with vertices of degree $\geq 2$.
If the size of the LR-core is $O(1)$, a simple brute force will be enough to 
further determine the MIS completely.
If, on the other hand, the LR-core has $O(N)$ size, this will take exponential time, 
implying that a simple LR algorithm fails.

\begin{figure}[h]
\includegraphics[width=8.5cm]{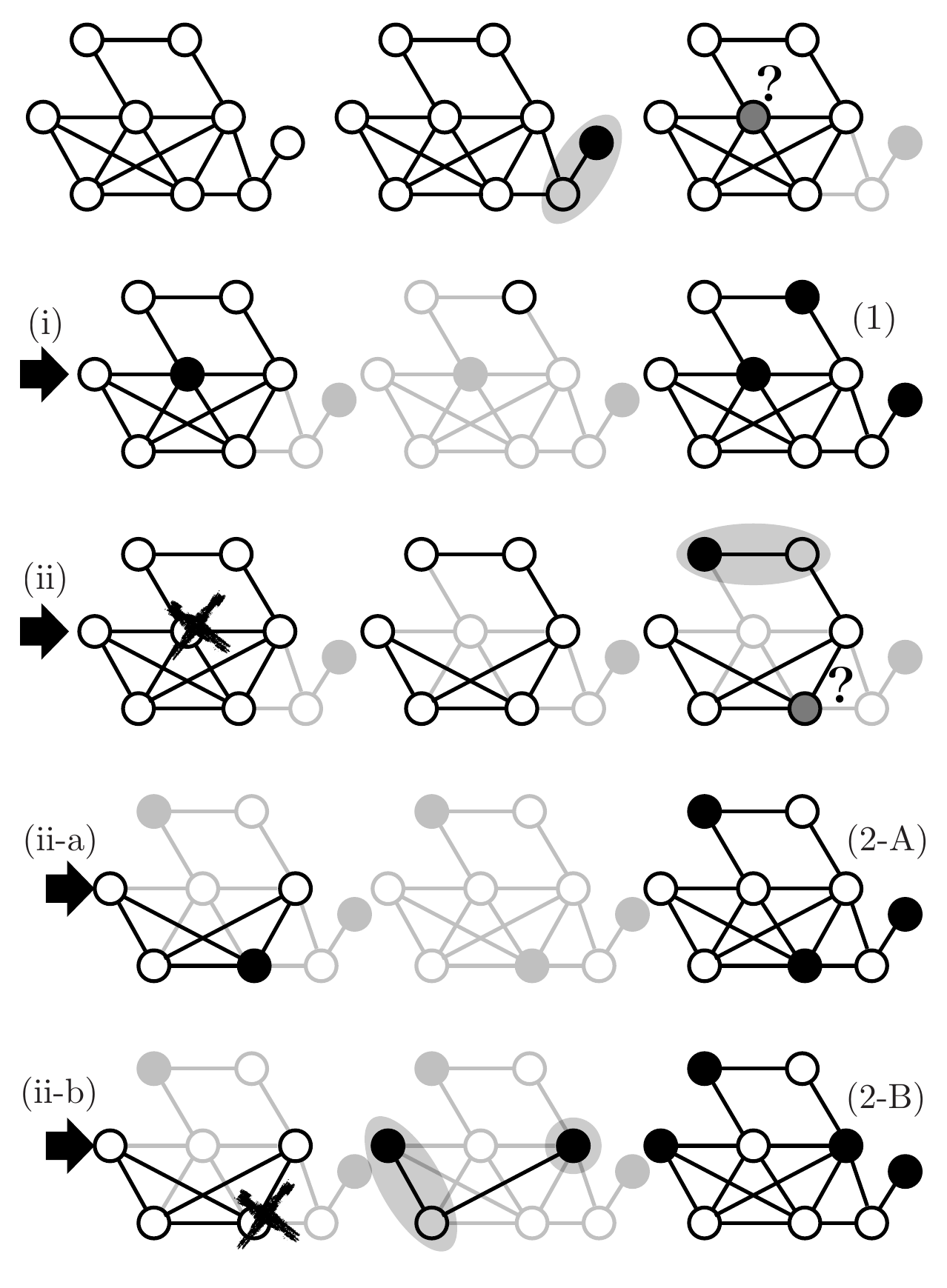}
\caption{\label{fig:DPLR} 
A schematic diagram for the dynamic programming leaf removal algorithm.
The gray-colored vertices are where the branching occurs,
represented by the arrows.
The cross mark means that the vertex is {\tt excluded} from the independent set.
The configuration and the size of the (so-far) maximum independent set found is 
always recorded, and is replaced whenever a larger independent set is found.
In this example, there are two branches (i,ii and a,b),
resulting in three different independent set configurations (1, 2-A, and 2-B).
The second one (2-A) has the same size as the first one (1), 
so a replacement will not occur until the third configuration (2-B) is found, 
which has the largest size.
 }
\end{figure}

Although we call any remaining subgraph as the LR-core indifferent to its statistical properties,
it is possible that some LR-cores are actually easier to attack.
For instance, it is possible that an LR-core is very fragile,
in the sense that if we remove one vertex from it, 
the remaining graph will actually become manageable with the LR algorithm.
If this is the case, the LR-core should not be considered as a fundamental obstacle.
Weather if an LR-core of some type is fragile or not is nontrivial.
In order to fully address the fragility of the LR-cores,
we introduce the Dynamic Programming Leaf Removal (DPLR) algorithm,
which is a combination of the LR algorithm with 
naive dynamic programming (DP),
 a common technique in computer science (Fig. \ref{fig:DPLR}) \cite{DynamicProgramming}.

The DPLR algorithm is simply the LR algorithm, whenever the graph does not have an LR-core.
Once when DPLR hits an LR-core, 
it chooses one vertex with the largest degree \cite{BrelazHeuristic}, and branches off for searching configurations 
{\tt including}/{\tt excluding} the selected vertex to the independent set.
It then starts the LR algorithm again, continuing this process until all the vertices are determined and the size of the independent set is calculated.
The algorithm then searches other branches left behind,
always remembering the largest-so-far independent set.
The algorithm is essentially conducting the perturbation we have argued above,
until the graph is completely turned into leaves.
In this way, we achieve a general protocol which reveals the fragility of the LR-core.
This algorithm was used for probing the hardness of a particular model in previous research \cite{UniqueMIS}.

\begin{figure}[h]
\includegraphics[width=8.5cm]{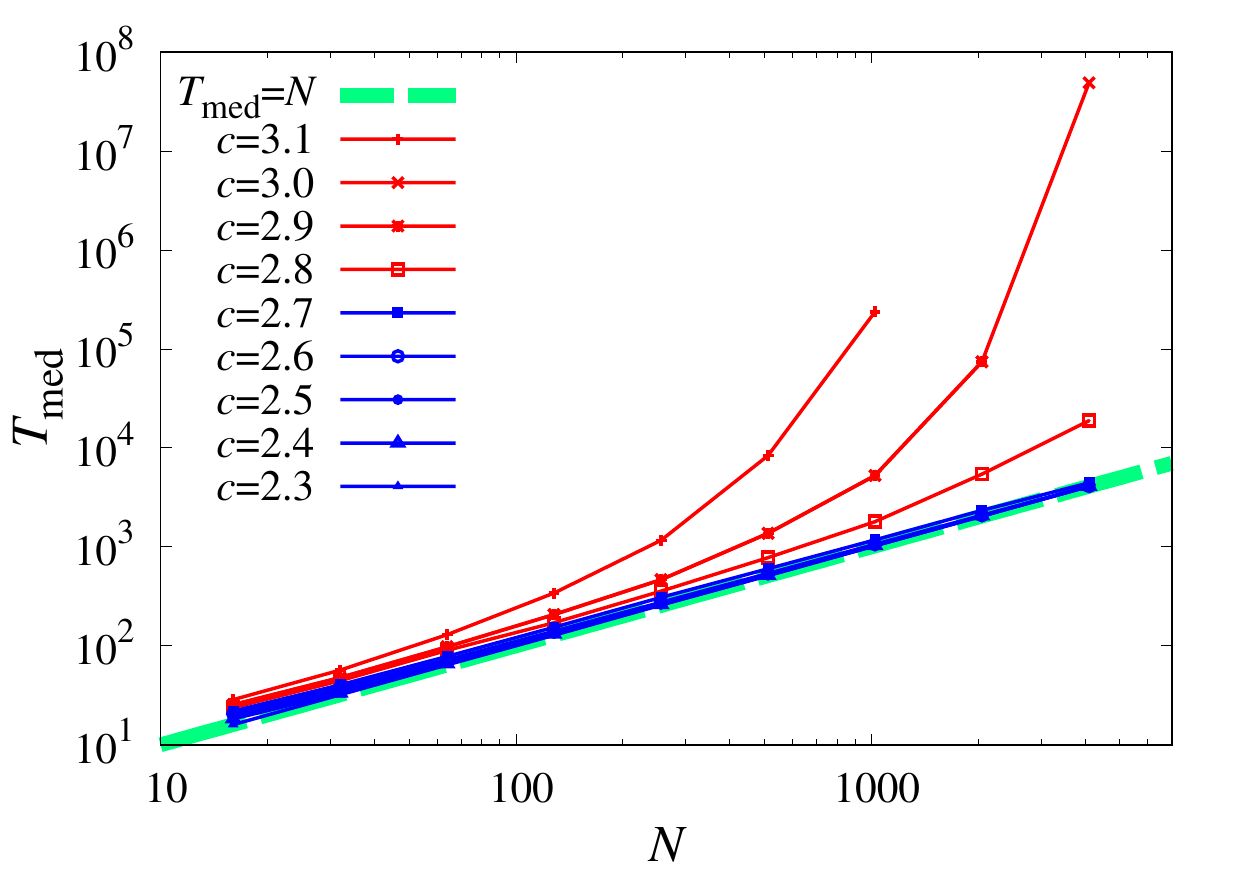}
\caption{\label{fig:ERmed} 
The median running time $T_{\mathrm{med}}$ of DPLR on 1024 Erd\"{o}s-R\'{e}nyi random graphs with different sizes and different average degree $c$.
 }
\end{figure}

\begin{figure}[h]
\includegraphics[width=8.5cm]{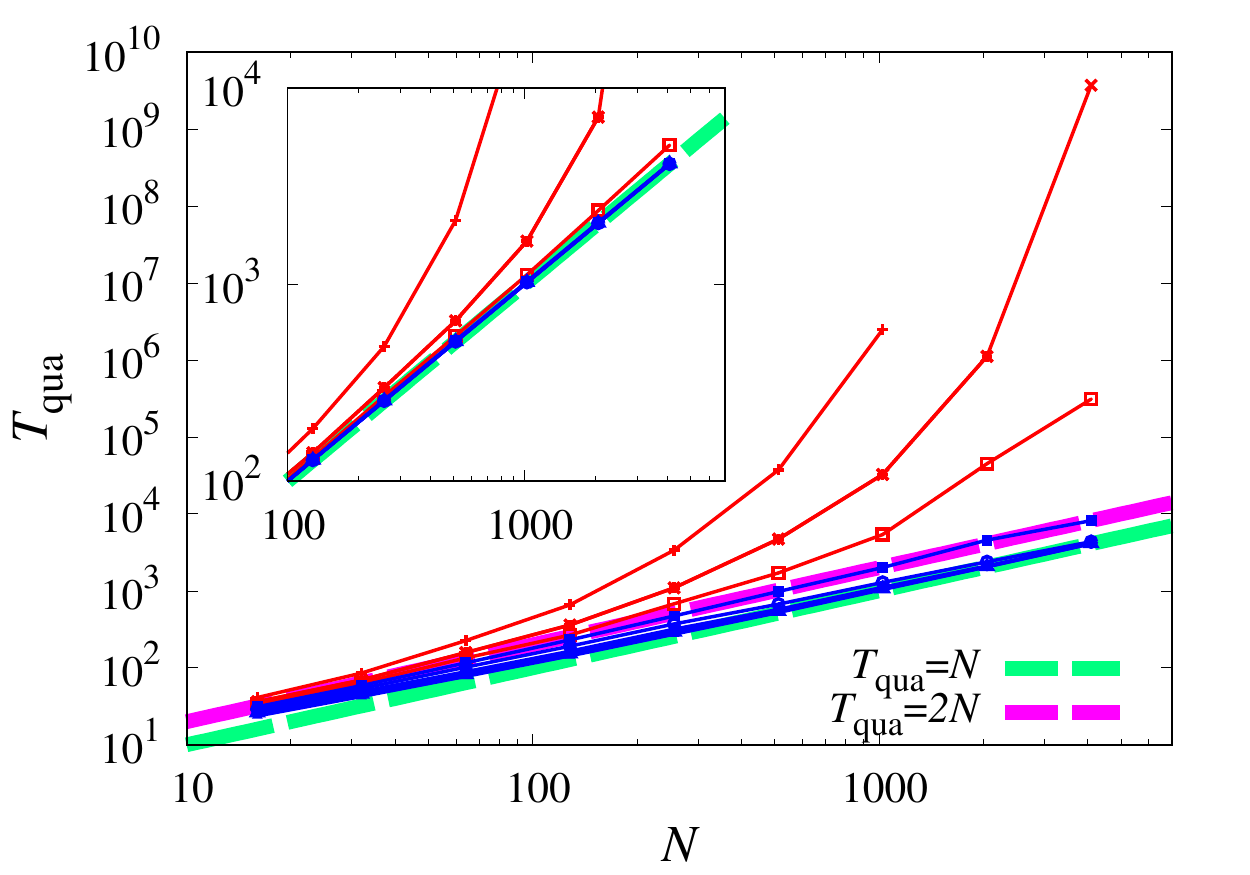}
\caption{\label{fig:ERqua} 
The first quartile running time $T_{\mathrm{qua}}$ of DPLR on 1024 Erd\"{o}s-R\'{e}nyi random graphs with different sizes and different average degree $c$.
Lines are expressed as the same as Fig. \ref{fig:ERmed}.
The inset shows the third quartile running time similarly.
 }
\end{figure}

Equipped with the DPLR algorithm, we are able to see the actual robustness of the LR-core.
We first discuss the simple Erd\"{o}s-R\'{e}nyi random graph.
Fig. \ref{fig:ERmed}
shows the median running time $T_{\mathrm{med}}$ of the DPLR algorithm 
on Erd\"{o}s-R\'{e}nyi random graph among 1024 samples.
The running time $T$ is defined by the number of decisions made for vertices to be {\tt included}/{\tt excluded} from the independent set.
All the lines for $c<e$ fall into the common line $T_{\mathrm{med}} = N$, which is the lower bound.
This means that at least half of the samples need no branching at all,
which is consistent with the fact that they lie in the region solvable by the simple LR algorithm.
On the other hand, we see convex curves for all $c>e$,
which implies super-polynomial growth of $T_{\mathrm{med}}$,
meaning that more than half of the LR-cores in all of the RSB region 
are actually very robust.
Fig. \ref{fig:ERqua} similarly shows the size dependence of the first (third for inset) quartile of 
the running time $T_{\mathrm{qua}}$ for different $c$.
No qualitative difference with Fig. \ref{fig:ERmed} is present,
which suggests that the typical behavior of random graphs is well-captured by the median value.
 The only quantitative difference is 
seen in the first quartile of $c=2.7$ 
where $T_{\mathrm{qua}}=2N$ 
which will be referred later.

\begin{figure}[h]
\includegraphics[width=8.5cm]{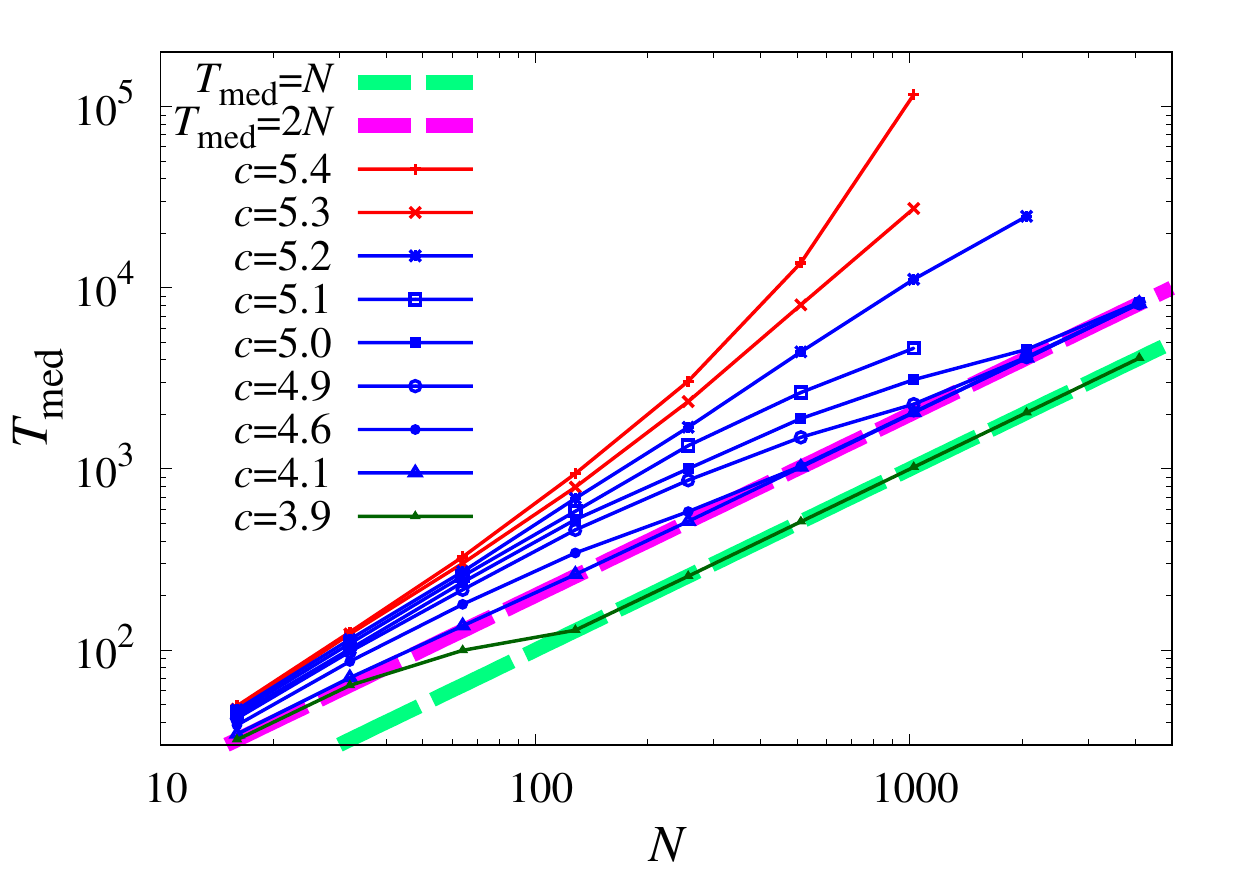}
\caption{\label{fig:SFmed} 
The median running time $T_{\mathrm{med}}$ of DPLR on 1024 scale-free CBA random graphs with different sizes and different average degree $c$.
 }
\end{figure}

\begin{figure}[h]
\includegraphics[width=8.5cm]{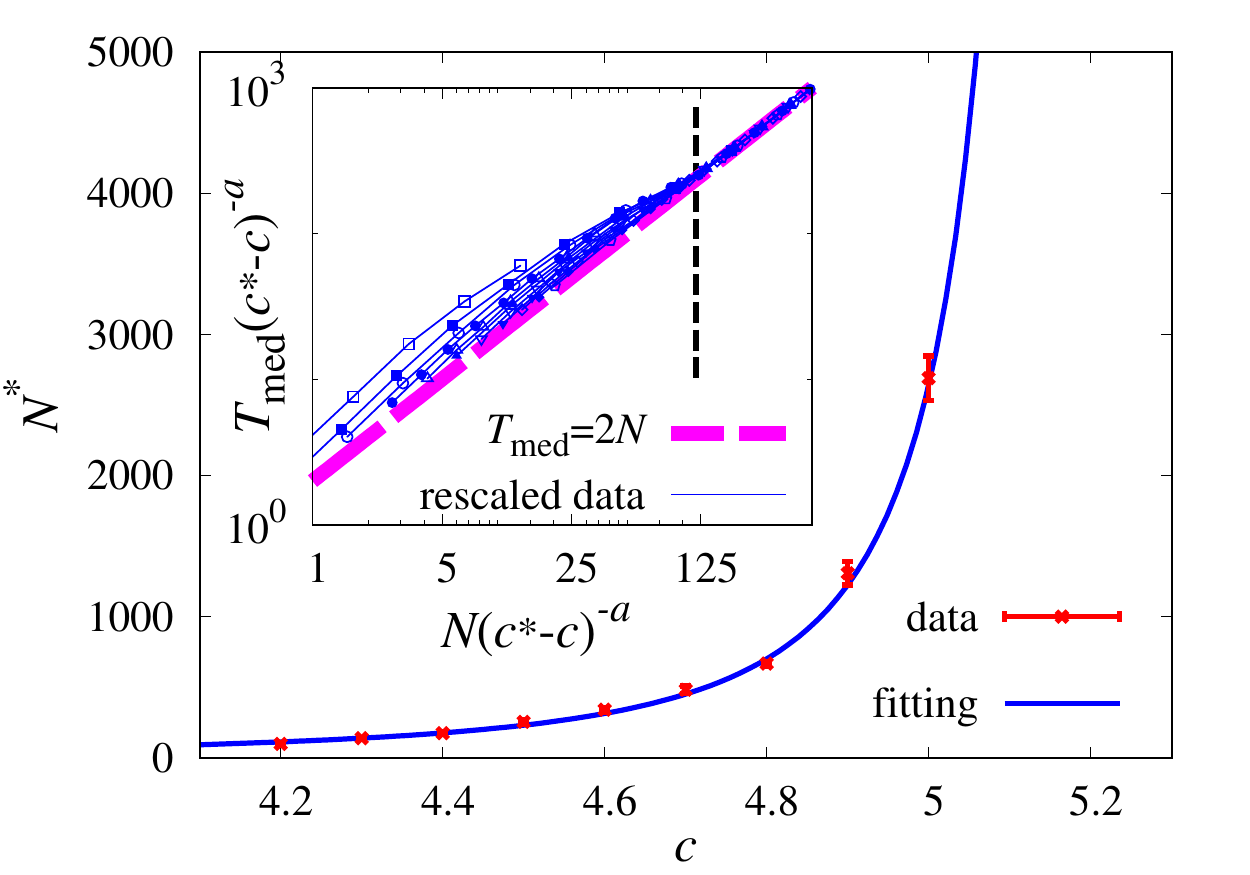}
\caption{\label{fig:SFFS} 
The $c$ dependence of the size $N^*$ where the finite size effect becomes small enough, 
i.e. when $T_{\mathrm{med}}/2N \leq 1.05$.
The error bars were evaluated by bootstrap.
The solid line indicates a result of least-squares fit to 
the form $N^*(c)=b(c^*-c)^a $, obtaining
$c^*=5.222 \pm 0.057
, b=117.8 \pm 11.5
$ and $a=-2.07 \pm 0.25$.
The inset shows the median running time divided by the rescaling factor $(c^*-c)^a$.
 All data sets ($c=4.2, 4.3, \ldots, 5.1$) seize to have finite size effects at a common point $N(c^*-c)^{-a}=b$,
shown by the vertical line.
 }
\end{figure}

DPLR and LR exhibited similar behaviors for Erd\"{o}s-R\'{e}nyi random graphs,
however the situation becomes different when we focus on scale free networks.
Fig. \ref{fig:SFmed} shows the median running time $T_{\mathrm{med}}$ of the DPLR algorithm 
on the CBA random graphs among 1024 samples.
Importantly, CBA graphs with $4<c<c_{\mathrm{RSB}}\simeq 5.239$ shows a linear growth of  $T_{\mathrm{med}}$.
This means the DPLR algorithm reduces the computation amount
compared to the naive LR algorithm in this parameter region,
from exponential to linear in $N$.
For small systems sizes we see finite size effects that makes the $T_{\mathrm{med}}$ larger than $2N$,
which seems as the asymptotic scaling.
Scale free networks have few vertices with very high degree,
which act as ``hubs".
These hubs make the LR-core fragile to DPLR-type perturbations, 
since they have many neighboring vertices which will be affected when 
deciding the hub to be {\tt included}/{\tt excluded} to the independent set.
Without these hubs, 
the graphs become robust against DPLR, which could be seen in the finite size effects.
The finite size effect is plotted in Fig. \ref{fig:SFFS}, which shows when the 
ratio $T_{\mathrm{med}}/2N$ becomes smaller than a certain value. 
The fitting suggests that the point where the finite size effect ends diverges at $c= 5.222 \pm 0.057$,
in well agreement with $c_{\mathrm{RSB}}\simeq5.239$.
Thus, although the
behavior for $c=5.2$ apparently seems nonlinear in Fig. \ref{fig:SFmed},
it is likely that 
it just has a long-lasting 
finite size effect  
until $N\sim 10^5$ as suggested by our scaling in Fig. \ref{fig:SFFS}.
Either way, the curves for $c>c_{\mathrm{RSB}}$ are convex where as
those of $c<c_{\mathrm{RSB}}$ are not,
meaning that the DPLR algorithm explicitly shows the RS-RSB/easy-hard correspondence.
We also see that the scaling of $T_{\mathrm{med}}$ changes from $N$ to $2N$ at $c_{\mathrm{LR}}=4$,
consistent with the emerging LR-core.
When $c_{\mathrm{LR}}<c$, all vertices have degree $\geq 2$,
forcing DPLR to branch at the very beginning.
Thus $T \geq 2N$, meaning that 
asymptotic scaling of $T_{\mathrm{med}}=2N$ implies
that the LR-core is {\it as fragile as is could possibly be}.
We believe that the situation is the same for $T_{\mathrm{qua}}$ at $c=2.7$ for the Erd\"{o}s-R\'{e}nyi graphs.

In conclusion, we have introduced a novel algorithm DPLR, 
which puts together the LR algorithm and DP.
We show explicitly that while the simple LR algorithm fails within the regime of 
$c_{\mathrm{LR}}<c<c_{\mathrm{RSB}}$,
DPLR moves in polynomial time (perhaps in linear time),
fully exhibiting the RS-RSB/easy-hard correspondence.
Our result has several important implications.

First, we should mention that not only the LR algorithm, but naive dynamic programming itself (branch-and-cut) takes exponential time in some parameter region within the RS phase \cite{binary}.
This means that although neither of the LR algorithm nor DP 
is enough to see the easy/hard transition induced by the RS/RSB transition, 
when put together, the DPLR algorithm reveals the true transition when the problem gets harder intrinsically.
This emphasizes the necessity of an adequate algorithm which
properly exhibits the RS-RSB/easy-hard correspondence.

Secondly, 
the scale free CBA model has a finite range in the parameter space which DPLR exhibits
asymptotic scaling of $T_{\mathrm{med}}=2N$, suggesting that 
``easiest possible LR-cores" occur quite naturally.
If $c=5.2$ indeed has an asymptotically linear scaling,
this means that the easy/hard transition is actually a linear-to-exponential transition,
more severe than polynomial-to-exponential.
Together with the fact that Erd\"{o}s-R\'{e}nyi model
surely has linear-to-exponential transition,
it is likely that ``moderately hard" LR-cores which take nonlinear polynomial time
are actually rare in random graphs.

Finally, we should emphasize that the agreement of the point which DPLR starts to take exponential time and the RS/RSB transition point is nontrivial.
The statistical mechanics analysis from which we obtain the RS/RSB transition does not rely on concepts such as the leaves.
They simply exploit the structure of the Hamiltonian describing the MIS problem.
The branching of DPLR,  on the other hand, takes advantage of the property of MIS 
which is that leaves are actually structures where the problem could be simplified.
The fact that these two different ways of analysis 
agree with each other on the phase transition point 
suggests the existence of RS-RSB/easy-hard correspondence.

\begin{acknowledgment}
We thank Y. Nishikawa for useful discussions. 
This research was supported by the Grants-in-Aid for Scientific Research
from the JSPS, Japan (No. 25120010). 
\end{acknowledgment}

\end{document}